\documentclass{pasj00}

\draft

\begin{document}
\SetRunningHead{S. Kato}{Frequency Correlations among Two-Armed p-mode Oscillations}
\Received{2012/00/00}
\Accepted{2012/00/00}

\title{An Atempt to Describe Frequency-Correlations among kHz QPOs and HBOs by Two-Armed 
Vertical p-mode Oscillations: Case of No Magnetic Field}

\author{Shoji \textsc{Kato}}
\affil{2-2-2 Shikanoda-Nishi, Ikoma-shi, Nara, 630-0114}
\email{kato@gmail.com, kato@kusastro.kyoto-u.ac.jp}

%


\maketitle

\begin{abstract}
Trapping of two-armed ($m=2$) vertical p-mode oscillations in relativistic disks are examined.
The disks are assumed to be isothermal in the vertical direction, but are
truncated at a certain height by the presence of corona.
The same issues have been examined in a previous paper (Kato 2012a).
In this paper, unlike the previous paper, however, we do not use the approximation that the oscillations are nearly vertical, 
but limit to a simpler case of no magnetic field.
As in the previous paper, the results suggest that the two basic oscillation modes 
[both are the fundamental ($n=1$) in the vertical direction
but in the horizontal direction one is the fundamental ($n_{\rm r}=0$) and the other 
the first overtone ($n_{\rm r}=1$)] correspond to the twin kHz QPOs.
Second, the oscillation mode which is the first overtone $(n=2)$ in the vertical direction and
the fundamental in the horizontal direction ($n_{\rm r}=0$) will correspond to the horizontal branch 
oscillation (HBO) of Z-sources.
The latter suggests that the horizontal branch of Z-sources is a sequence of temperature change
in disks whose vertical thickness is strongly terminated.
The temperature increases leftward along the sequence from the apex between normal and horizotal branches.

\end{abstract}

\section{Introduction}

KHz quasi-periodic oscillations (kHz QPOs) observed in neutron-star low-mass X-ray binaries (NS LMXBs)
will give an important clue to understand the innermost structure of relativistic
disks of NS LMXBs, if their origin is clarified.
So far, many models of kHz QPOs are proposed.
Among them trapped disk oscillations are one of promising candidates, and trapping of p-, g-, and
c- mode oscillations in relativistic disks has been extensively studied
(see Wagoner 1999, Kato 2001, and Kato et al. 2008 for review, and see Fu \& Lai 2009 and Lai \& Tiang 2009
for related topics).
  
In addition to the above trapped oscillations, Kato (2010, 2011a,b, 2012a,b) suggested that 
trapped, two-armed ($m=2$), vertical p-mode
oscillations in geometrically thin (magnetized) disks are one of possible 
origins of kHz QPOs observed in neutron-star X-ray sources.
In this model two trapped oscillations, both of which are fundamental modes ($n=1$) in the vertical
direction but one is the fundamental ($n_r=0$) and the other the first overtone ($n_r=1$) in the radial 
direction, are regarded to be the twin kHz QPOs.
Frequencies of these two oscillations change with correlation by time changes of disk temperature, 
disk vertical thickness and toroidal magnetic fields in disks.
This correlated change of the frequencies is considered to be the cause of the
observed correlated change of the twin kHz QPOs (Kato 2011b, 2012b).

In calculating the frequencies of trapped, two-armed p-mode oscillations, we have assumed that the
disks are isothermal in the vertical direction and have toroidal magnetic fields.
Furthermore, for simplicity, the toroidal magnetic fields are assumed to be distributed in the vertical 
direction so that the Alfv{\' e}n speed, $c_{\rm A}$, is constant in the vertical direction
(Kato 2011a, 2012a).
Even in this simplified case, semi-analytical calculations of frequencies of 
trapped vertical p-mode oscillation are still complicated.
Hence, in previous studies (Kato 2011a, 2012a) we have assumed that the oscillations are nearly  
vertical and the motions in the radial
direction associated with the oscillations can be treated as small perturbations over the vertical motions.
The approximation is allowed in qualitative arguments, but not accurate enough 
to do quantitative comparison of our model with observations.

Hence, in this paper, returning to a simpler case, i.e., to the case of non-magnetized disks, 
we examine the trapping of two-armed vertical p-mode oscillations, since in this simpler case
the frequencies of trapped oscillations can be calculated analytically without introducing the 
approximation that the oscillations are nearly vertical.
The results show that even in this case of non-magnetized disks, the observed correlation of twin
kHz QPOs and the correlation between kHz QPOs and horizontal branch oscillation can be both described
by moderate time variations of disk temperature, if the disks  
are strongly truncated in the vertical thickness.

\section{Disk Models and Parameters Describing the Disks}

Distinct from previous papers by Kato (2011a,b, 2012a,b), no magnetic field is assumed in this paper.
Except for this, our disk models are the same as those adopted in previous papers.
That is, the unperturbed axisymmetric disks are vertically isothermal.
Thus, the vertical hydrostatic equilibrium gives that in the vertical direction the density $\rho_0(r,z)$ is
distributed as
\begin{equation}
       \rho_0(r,z)=\rho_0(r){\rm exp}\biggr(-\frac{z^2}{2H^2}\biggr),
\label{2.1}
\end{equation}
where the scale height $H(r)$ is related to the isothermal accoustic speed $c_{\rm s}(r)$ and the vertical
epicyclic frequency $\Omega_\bot(r)$ by
\begin{equation}
     H^2(r)=\frac{c_{\rm s}^2}{\Omega_\bot^2(r)}.
\label{2.2}
\end{equation}
Here, $r$ is the radial coordinate of the cylindrical coordinates ($r$,$\varphi$,$z$) whose
$z$-axis is perpendicular to the unperturbed disk plane and the origin is at the disk center.
It is noted that the Newtonian formulation is adopted in this paper, except when the 
radial distributions of $\Omega(r)$, $\Omega_\bot(r)$, and $\kappa(r)$ are considered, where
$\kappa(r)$ is the radial epicyclic frequency and $\Omega(r)$ is the angular velocity of disk rotation 
and taken to be the Keplerian angular
velocity, $\Omega_{\rm K}(r)$, since geometrically thin disks are considered.

We assume that the isothermal disks are terminated at a certain height and surrounded by hot, 
low-density corona.
The height of termination, $z_{\rm s}(r)$, is taken to be a free parameter.
Hereafter, we introduce a dimensionless parameter, $\eta_{\rm s}$, defined by $\eta_{\rm s}\equiv
z_{\rm s}/H$, which is a function of $r$ in general, but taken to be a constant independent of $r$, 
since the trapped region of oscillations is not wide, except for the case of small $\eta_{\rm s}$, 
as is shown below (see figures 2 and 3).

The disk temperature is also a parameter describing disks.
As a reference, we adopt the temperature in the standard disks where the gas pressure dominates over 
the radiation pressure and the opacity comes from the free-free processes.
In this case, if the conventional viscosity parameter $\alpha$ is 0.1 and the mass accretion rate 
normalized by the Eddington critical accretion rate is 0.3,
the square of the isothermal accoustic speed, $(c_{\rm s}^2(r))_0$, is given by (e.g., Kato et al. 2008)
\begin{equation}
   (c_{\rm s}^2(r))_0=1.79\times 10^{16}\biggr(\frac{M}{M_\odot}\biggr)^{-1/5}
      \biggr(\frac{r}{r_{\rm g}}\biggr)^{-9/10} \ \ {\rm cm}^2\ {\rm s}^{-2},
\label{2.3}
\end{equation}
where $M$ is the mass of the central source and $r_{\rm g}$ is the Schwarzschild radius defined by
$r_{\rm g}=2GM/c^2$.
In this paper we adopt $\beta[\equiv c_{\rm s}^2/(c_{\rm s}^2)_0]$ as a dimensionless parameter
describing the temperature distribution in disks.
$\beta$ is, in general, a function of $r$, but taken to be constant independent of $r$, since
the trapped region of oscillations is generally not wide, as mentioned before.

In summary, we specify our disk models by two parameters:
\begin{equation}
  \beta\equiv \frac{c_{\rm s}^2}{(c_{\rm s}^2)_0}\quad {\rm and}\quad
   \eta_{\rm s}\equiv \frac{z_{\rm s}}{H}
\label{2.4}
\end{equation}
and two additional parameters $M/M_\odot$ and $a_*$, where $a_*$ is the spin parameter
of the central source ($0\leq a_*<1.0$).

\section{Trapped Oscillations and Their Frequencies}

Wave equations describing isothermal oscillations in vertically isothermal and geometrically thin disks 
have been well studied (e.g., Kato 2001 for review).
It is convenient to introduce the variable $h_1$ defined by $h_1=p_1/\rho_0$, where 
$p_1$ is the pressure variation over the unperturbed pressure $p_0$ and $\rho_0$ is the
unperturbed density.
Then, under the approximation that the radial wavelength of the perturbations is smaller
than the scale length of the radial variations of the equilibrium properties of the disk, 
we have a wave equation in terms of $h_1$ as (e.g., Kato 2001)
\begin{equation}
    \frac{1}{\rho_0}\frac{\partial}{\partial z}\biggr(\rho_0\frac{\partial h_1}{\partial z}\biggr)
    +(\omega-m\Omega)^2\frac{\partial}{\partial r}\biggr[\frac{1}
        {(\omega-m\Omega)^2-\kappa^2}\frac{\partial h_1}{\partial r}\biggr]
        +\frac{(\omega-m\Omega)^2}{c_{\rm s}^2}h_1=0,
\label{3.1}
\end{equation}
where the perturbations are taken to be proportional to exp$[i(\omega t-m\varphi)]$,
$\omega$ and $m$ are being, respectively, the frequency and the azimuthal wavenumber of perturbations.
In this paper we consider only the case of $m=2$, but $m$ is retained hereafter without specifying $m=2$
so that the origin of $m$ can be traced back.
Let us introduce a dimensionless variable $\eta$ defined by $\eta\equiv z/H$. 
Then, equation (\ref{3.1}) is written as
\begin{eqnarray}
   &&\frac{d^2h_1}{d\eta^2}-\eta\frac{dh_1}{d\eta}
        +H^2(\omega-m\Omega)^2\frac{\partial}{\partial r}
         \biggr[\frac{1}{(\omega-m\Omega)^2-\kappa^2}
                \frac{\partial h_1}{\partial r}\biggr] \nonumber \\
   &&     +\frac{(\omega-m\Omega)^2}{\Omega_\bot^2}h_1=0.
\label{3.2}
\end{eqnarray}

Wave equations similar to or more general than equation  (\ref{3.2}) have been solved by the WKB method, 
approximately separating $h_1(r,\eta)$ as $h_1(r,\eta)=g(\eta)f(r, \eta)$
(e.g., Silbergleit et al 2001, Ortega-Rodrigues et al. 2008, Kato 2012a).
In the present case, $h_1$ is simply separated as $h_1(r,\eta)=g(\eta)f(r)$, and we have a set of two
equations (e.g., Okazaki et al. 1987):
\begin{equation}
      \frac{d^2g}{d\eta^2}-\eta\frac{dg}{d\eta} +Kg=0,
\label{3.3}
\end{equation}
and
\begin{equation}
   c_{\rm s}^2\frac{d}{dr}\biggr[\frac{1}{(\omega-m\Omega)^2-\kappa^2}\frac{df}{dr}\biggr]
   +\biggr[1-\frac{K\Omega_\bot^2}{(\omega-m\Omega)^2}\biggr]f=0,
\label{3.4}
\end{equation}
where $K$ is the separation constant and can be obtained by solving equation (\ref{3.3}) with
relevant boundary conditions, which is given in the next subsection.

\subsection{Oscillations in the Vertical Direction and Their Eigen-Frequencies}

In the case where the isothermal disks extend infinity in the vertical direction (i.e., $\eta_{\rm s}=\infty$),
the boundary condition of finite wave-energy density at $\eta=\pm\infty$
requires that $K$ is zero or a positive integer, say $n$, and $g(\eta)$ is a Hermite polynomial, 
${\cal H}_n(\eta)$ (Okazaki et al.1987).
The integer $n$ thus represents the number of nodes of $h_1$ in the vertical direction.
In the present problem, we are interested in vertical p-mode oscillations
and thus $n$ starts from $n=1$, not from $n=0$ ($n=0$ corresponds to the horizontal p-mode oscillations and
the oscillations of $n=0$ is outside of our interest in this paper.)
Hereafter, we use $n(=1,2,3...)$ to denote node number of $h_1$ in the vertical direction.

In the case of vertically terminated disks ($\eta_{\rm s}\not=\infty$), we impose boundary conditions
that the Lagrangian pressure variation vanishes at $\eta=\pm\eta_{\rm s}$, i.e., $\delta p_1=0$
at $\eta=\pm\eta_{\rm s}$, since in the coronae the
perturbations propagate away rapidly because of high temperature.
If we adopt the approximation that the oscillations are nearly vertical there, this boundary
condition can be reduced to (Kato 2012a)
\begin{equation}
      \frac{\partial u_z}{\partial z}=0 \quad {\rm at} \quad \eta=\pm \eta_{\rm s},
\end{equation}
where $u_z$ is the $z$-component of velocity associated with the oscillations.
Since the $z$-component of equation of motion is related to $h_1$ by
$i(\omega-m\Omega)u_z=-\partial h_1/\partial z$,
this boundary condition can be expressed in terms of $g$ as
\begin{equation}
       \frac{d^2g}{d\eta^2}=0 \quad {\rm at}\quad \eta=\pm\eta_{\rm s}.
\end{equation}
The eigenvalue $K$ obtained by solving equation (\ref{3.3}) with this boundary condition
is expressed hereafter by ${\tilde K}_{n, {\rm s}}$, since it depends upon $\eta_{\rm s}$ and
the number of node, $n$.
The tilde is attached to $K$ in order to avoid confusion with $K_{n,{\rm s}}$ in Kato (2012a).
The results are summarized in table 1 for some values of $n$ and $\eta_{\rm s}$.
The eigenvalue ${\tilde K}_{n,{\rm s}}$ is related to $K_{n,s}$ obtained by the boundary condition
$dg/d\eta=0$ by ${\tilde K}_{n,{\rm s}}= 1+K_{n,{\rm s}}$.\footnote{
Let us denote the eigenvalue and eigenfinction of equation:
$$ \frac{d^2g}{d\eta^2}-\eta\frac{dg}{d\eta}+Kg=0,   \eqno{(*)}
$$
$K_{n,{\rm s}}$ and $g^u$, respectively, when the equation is solved by the boundary condition
$dg/d\eta=0$.
Then, $K_{n,{\rm s}}$ and $g^u$ are related to ${\tilde K}_{n,{\rm s}}$ and $g$ given in this paper
by $K_{n,{\rm s}}={\tilde K}_{n,{\rm s}}-1$ and $g^u=dg/\eta$, respectively.
This can be shown by substituting $g^u$ given above into equation (*) and using the fact that
$g$ is a solution of equation (*).
}

\begin{longtable}{cccccccc}
\caption{Eigenvalue ${\tilde K}_{n,{\rm s}}$ of Vertical Oscillations}
\label{table 1}
\endfirsthead
\hline\hline

  node number & \multicolumn{7}{c} {termination height $\eta_{\rm s}$} \\

\hline
        & 1.0 & 1.3 & 1.5 & 1.8 & 2.0 & 2.5 & $\infty$ \\
\hline
  $n=1$ & 1.0 & 1.0 & 1.0 & 1.0 & 1.0 & 1.0 & 1.0      \\
  $n=2$ & 4.00& 3.016 & 2.670 & 2.365 & 2.243 & 2.083 & 2.0      \\ 

\hline
\end{longtable}

\subsection{Frequencies of Trapped Oscillations}

We solve equation (\ref{3.4}) by the WKB method with ${\tilde K}_{n,{\rm s}}$ given above.
By introducing $\tau(r)$ defined by, as Silbergleit et al. (2001) did,
\begin{equation}
     \tau(r)=\int_{r_{\rm i}}^r\biggr \{[\omega-m\Omega(r')]^2-\kappa^2(r'))\biggr\}dr',
\label{3.5}
\end{equation}
where $r_{\rm i}$ is the radius where inner boundary condition is imposed, we can reduce 
equation (\ref{3.4}) in the form
\begin{equation}
     \frac{d^2f}{d\tau^2}+Qf=0,
\label{3.6}
\end{equation}
where
\begin{equation}
     Q(\tau)=\frac{1-{\tilde K}_{n,{\rm s}}\Omega_\bot^2/(\omega-m\Omega)^2}
              {c_{\rm s}^2[(\omega-m\Omega)^2-\kappa^2]}.
\label{3.7}
\end{equation}
As $r_{\rm i}$ we take the radius of the marginally stable circular orbit 
and assume that $h_1$ vanishes there, i.e, $f=0$.\footnote{
The present boundary condition, $f=0$, is consistent with the boundary condition adopted by Kato (2012a), 
which is $u_z=0$ at $r=r_{\rm i}$.
This is because the $z$-component of equation of motion  
is $i(\omega-m\Omega)u_z=-\partial h_1/\partial z$ 
and since $\partial h_1/\partial z=0$ when $h_1$ 
is separated as $h_1=g(\eta)f(r)$ and $f=0$.
}
The other boundary condition is that outside of the capture radius, $r_{\rm c}$, 
which is defined by $Q(r_{\rm c})=0$,
the oscillations are spatially damped.
Then, as the trapping condition, we have (see Silbergleit et al. 2001; Kato 2012a)
\begin{equation}
   \int_{r_{\rm i}}^{r_{\rm c}} Q^{1/2}(\tau)d\tau
    =\int_{r_{\rm i}}^{r_{\rm c}} \frac{[(\omega-m\Omega)^2-\kappa^2]^{1/2}}{c_{\rm s}}
    \biggr[1-\frac{{\tilde K}_{n,{\rm s}}\Omega_\bot^2}{(\omega-m\Omega)^2}\biggr]^{1/2}dr
    = \pi(n_{\rm r}+3/4),
\label{3.8}
\end{equation}
where $n_{\rm r}(=0,1,2,...)$ is the node number of $f$ in the radial region of 
$r_{\rm i} < r < r_{\rm s}$.
This equation gives the frequencies of trapped oscillations (see also the appendix),
as functions of $n$, $n_{\rm r}$, $\beta$, 
and $\eta_{\rm s}$ for given $M$ and $a_*$.

\section{Numerical Results on Trapped Oscillations}

Numerical results obtained by solving equation (\ref{3.8}) are shown here mainly in the case of 
$M=1.4M_\odot$ with some values of $a_*$, $\beta$, and $\eta_{\rm s}$.
Among many oscillation modes of $n(=1,2,3,...)$ and $n_{\rm r}(=0,1,2,...)$, our main interests here are on
oscillations modes of $(n,n_{\rm r})=(1,0)$, (1,1), and (2,0),  since we think that the former two will 
correspond to the set of twin kHz QPOs, and the third one will to the horizontal branch oscillation (HBO).

\begin{figure}
\begin{center}
    \FigureFile(80mm,80mm){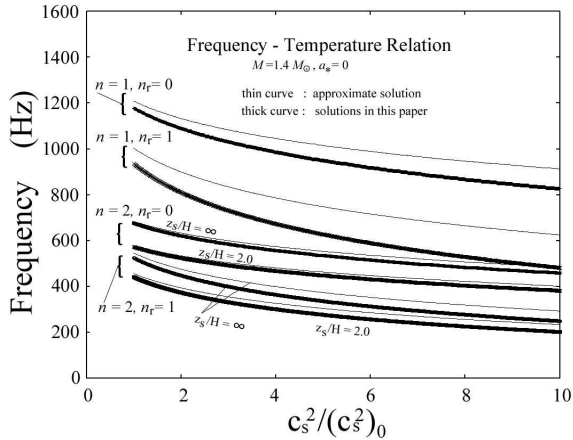}
\end{center}
\caption{Frequency -- temperature relation of trapped two-armed ($m=2$) vertical p-mode oscillations
in isothermal disks.
Two cases are shown where the disks extend infinitely in the vertical direction $(z_{\rm s}/H=\infty$)
and are terminated at  the height of $2H$, i.e., $\eta_{\rm s}\equiv z_{\rm s}/H=2.0$.
Four oscillation modes are shown, i.e., the set of ($n$, $n_{\rm r}$) are
(1,0), (1,1), (2,0) and (2,1).
The thin curves are for the cases where the frequencies are calculated with the approximation that
the horizontal motions associated with the oscillations are small perturbations over the vertical ones.
In oscillations of $n=1$, the frequencies are independent of $\eta_{\rm s}$.
The mass $M$ and the spin parameter $a_*$ adopted are $M=1.4M_\odot$ and $a_*=0$.
 }
\end{figure}

Figure 1 shows the $\beta$-dependence of frequencies of the trapped oscillations.
Four oscillation modes of $(n,n_{\rm r})=(1,0)$, (1,1), (2,0), and (2,1) are shown for two cases of 
$\eta_{\rm s}=\infty$ and $\eta_{\rm s}=2.0$ with $a_*=0$.
For comparison, the results obtained by the approximation method that the horizontal motions associated with
the oscillations are small perturbations over the vertical ones (Kato 2010, 2011a) are shown by thin curves.
It is noted that in the case of $n=1$, the frequencies are independent of $\eta_{\rm s}$, 
since ${\tilde K}_{n,{\rm s}}$ is 1.0 and free from $\eta_{\rm s}$.
Frequencies of oscillations decrease in the order of $(n,n_{\rm r})=(1,0)$, (1,1), (2,0), and (2,1), 
as long as $\beta<10.0$, and decrease as $\beta$ increases.
It is also noted that the frequencies derived by using the approximation of nearly vertical oscillations
are systematically higher than those derived without the approximation, especially
in the case of $n=1$.

\begin{figure}
\begin{center}
    \FigureFile(80mm,80mm){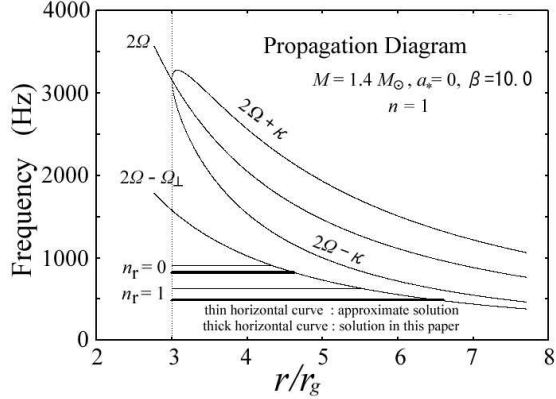}
\end{center}
\caption{Propagation diagram for two oscillations of ($n$, $n_{\rm r})$ = (1,0) and (1,1).
The horizontal lines show the radial range where oscillations are trapped.
The thin lines are for the cases where the approximation of nerly vertical oscillations is adopted
to calculate frequencies.
$\beta\equiv c_{\rm s}^2/(c_{\rm s}^2)_0=10.0$ has been adopted.
As  in figure 1, $M=1.4M_\odot$ and $a_*=0$ are taken.}
\end{figure}

Figure 2 is the propagation diagram for two oscillations of $(n,n_{\rm r})=(1,0)$ and (1,1) 
in the case of $\beta=10.0$.
Disk parameters adopted in figure 2 are the same as those in figure 1.
For comparison, the oscillations derived by the approximation of nearly vertical oscillations
are shown by thin curves.

To understand the propagation diagram of figure 2, let us examine propagation and evanescent regions of 
an oscillation with frequency $\omega_0$.
Since the propagation regions are specified by $Q>0$ [see equation (\ref{3.6})], 
we know that the regions are $(\omega_0-m\Omega)^2-{\tilde K}_{n,{\rm s}}\Omega_\bot^2>0$ and
$(\omega_0-m\Omega)^2-\kappa^2 <0$ in the case where ${\tilde K}_{n,{\rm s}}\geq 1$.
The former is the propagation region of vertical p-mode oscillations and separated into two 
regions of $\omega_0-m\Omega<-{\tilde K}_{n,{\rm s}}^{1/2}\Omega_\bot$
and $\omega_0-m\Omega>{\tilde K}_{n,{\rm s}}^{1/2}\Omega_\bot$, and
the latter is the propagation region of g-mode oscillations and $-\kappa<\omega_0-m\Omega<\kappa$.
In the case of vertical p-mode oscillations with $n=1$, we have ${\tilde K}_{n,{\rm s}}=1$.
Hence, for oscillations of $m=2$, $\omega_0<2\Omega-\Omega_\bot$ is  
a propagation region, which is the region of our interest in this paper.
The radius where $\omega_0=2\Omega-\Omega_\bot$ is the capture radius, $r_{\rm c}$.
Outside of $r_{\rm c}$, there is an evanescent region which extends till the radius of $\omega_0=2\Omega-\kappa$
(inner Lindblad radius).
The region specified by $2\Omega-\kappa<\omega_o<2\Omega+\kappa$ is again a propagation region, but
the vertical p-mode oscillations which are nearly trapped in the region of $\omega_0<2\Omega-\Omega_\bot$
will penetrate only weakly to this propagation region. 
It is noted that inside the region of $2\Omega-\kappa<\omega_o<2\Omega+\kappa$ there is the corotation point, 
$\omega_0=2\Omega$, where the oscillations will be damped.
In the present oscillations, however, their damping at the corotation resonance will be negligible, 
since wave leakage to the point will be negligible by spatial damping in the evanescent region,
as mentioned above.
The trapping condition (\ref{3.8}) determines the frequency and the capture radius which are shown in figure 2
by thick horizontal lines.
The thin horizontal lines show the case where the oscillations are approximated to be nearly 
vertical ones.

\begin{figure}
\begin{center}
    \FigureFile(80mm,80mm){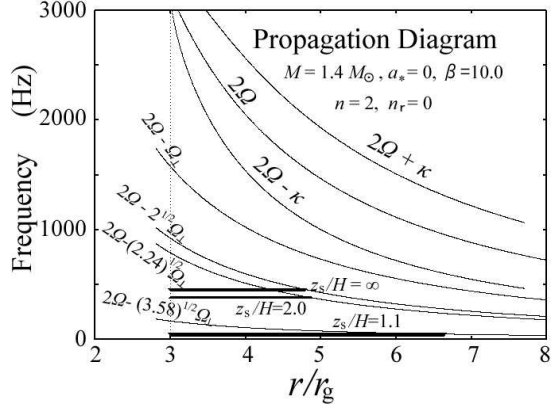}
\end{center}
\caption{Propagation diagram for oscillations of ($n$, $n_{\rm r})$=(2,0) for three cases of 
$\eta_{\rm s}=\infty$, 2.0, and 1.1.
Other parameters are the same as figure 2.  }
\end{figure}

Figure 3 shows the propagation diagram of oscillations of $n=2$ and $n_{\rm r}=0$ in the case of 
$\beta=10.0$. Three cases of $\eta_{\rm s}$ (i.e., $\eta_{\rm s}=\infty$, $2.0$, and 1.3) 
are shown with other parameters being the same as those in figures 1 and 2.
A purpose of showing this figure is to demonstrate that in the case of $n=2$ the curve specifying 
the capture radius, i.e., the radius of $\omega_0=2\Omega-{\tilde K}_{n,{\rm s}}^{1/2}\Omega_\bot$ depends on 
$\eta_{\rm s}$ since ${\tilde K}_{n,{\rm s}}$ depends on $\eta_{\rm s}$.
It is notices that when $\eta_{\rm s}$ is close to 1,  the trapped region is rather wide, since
${\tilde K}_{n,{\rm s}}$ is large and close to 4.0.

\begin{figure}
\begin{center}
    \FigureFile(80mm,80mm){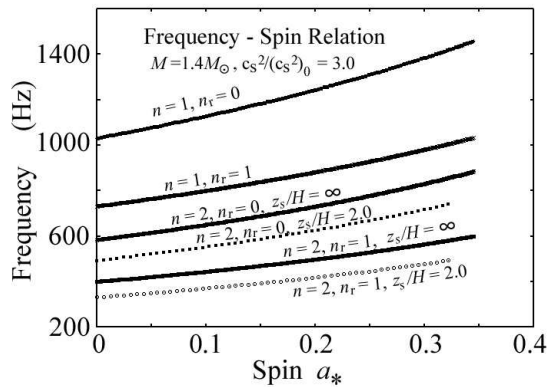}
\end{center}
\caption{Frequency -- spin relation for four oscillation modes of ($n$, $n_{\rm r})=$ (1,0), (1,1), (2,0),
and (2,1).
In the case of $n=2$, the relation depends on $\eta_{\rm s}$.
Two cases of $\eta_{\rm s}=\infty$ and 2.0 are shown for the oscillation of $n=2$.
$M$ and $\beta$ adopted are $M=1.4M_\odot$ and $\beta=3.0$.  }
\end{figure}

Figure 4 shows the frequency-spin relation for four oscillation modes of 
$(n,n_{\rm r})=(1,0)$, (1,1), (2,0), and (2,1) in the case of $\beta=10.0$ with $M=1.4M_\odot$.
Two cases of $\eta_{\rm s}=\infty$ and $\eta_{\rm s}=2.0$ are shown.
As $a_*$ increases the frequencies of trapped oscillations increase, since the oscillations are
more localized in the inner region as $a_*$ increases.

\section{Frequency Correlation and Comparison with Oscillations}

We take the standpoint that two oscillations of $(n, n_{\rm r})=$ (1,0) and (1,1) with $m=2$ are, respectively, 
the upper and lower kHz QPOs.
In addition, we suppose that the (2,0) oscillation corresponds to the horizontal branch oscillation
(HBO).
These possibilities are examined here by comparing correlation curves derived by the model with observations.

\subsection{Frequency Correlation between (1,0) and (1,1) Oscillations}

We examine how the set of (1,0) and (1,1) oscillations move on a 
frequency-frequency diagram when $\beta$ is changed from 1.0 to 10.0.
First, two cases of $a_*=0$ and 0.2 are shown in figure 5 with $M=1.4M_\odot$.
These correlation curves are superposed on the diagram showing the observed
frequency-frequency plots of the twin kHz QPOs.
It should be noted here that the correlation curves are free from $\eta_{\rm s}$, since
the frequencies of the $n=1$ oscillations are independent of $\eta_{\rm s}$.
An increase of $a_*$ shifts the correlation curve in the upper-right direction.
Let us next consider effects of mass of the central source on the correlation curve.
Figure 6 is for $M=1.8M_\odot$ and three cases of $a_*=0$, 0.2, and 0.4 are shown.
This figure shows that an increase of mass shifts the correlation curve in the lower-left direction
on the diagram, compared with the corresponding cases of $M=1.4M_\odot$. 

\begin{figure}
\begin{center}
    \FigureFile(80mm,80mm){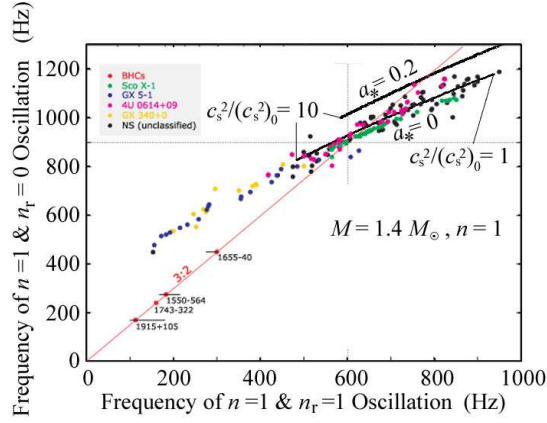}
\end{center}
\caption{The frequency correlation between ($n$, $n_{\rm r}$)$=$ (1,0) and (1,1) oscillations
for two cases of $a_*=0$ and 0.2.
The value of $\beta$ is changed from 1.0 to 10.0.
$M=1.4M_\odot$ is adopted.
The plot of observational data of some typical Z-sources, taken from the figure of 
Abramowicz (2005), are overlapped on this figure.
The straight line labelled by 3 : 2 is the line on which frequency ratio is 3 : 2.  (Color Online) }
\end{figure}

\begin{figure}
\begin{center}
    \FigureFile(80mm,80mm){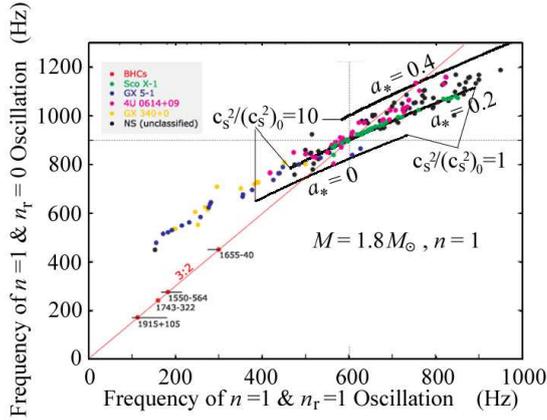}
\end{center}
\caption{The same as figure 5, except that $M=1.8M_\odot$ and three cases of
$a_*=0$, 0.2, and 0.4 are considered.  (Color Online) }
\end{figure}

From figures 5 and 6 we see that the observed correlation between 
the twin kHz QPOs of typical Z-sources can be well described by $a_*\sim 0$ if the mass of the central source is 
slightly larger than $1.4M_\odot$.
A larger $a_*$, say $a_*\sim 0.2$, is required if the mass is larger, say $1.8 M_\odot$.


In previous papers (Kato 2011a,b, 2012a,b), we have calculated frequencies of trapped vertical p-mode 
oscillations and correlations among them in the case where the disks have toroidal magnetic fields.
In these papers, however, we adopted the approximation that the oscillations are nearly vertical in the sense 
that the horizontal motions associated with the oscillations can be treated as small perturbations on 
the vertical motions.
This approximation is qualitatively acceptable, but not always relevant to quantitative arguments.
In order to see this more in detail, we show here the results derived by
using the approximation in the particular case  of no magnetic fields, since in this case we have derived the 
frequencies without the approximation.
The correlation curves of the (1,0) and (1,1) oscillations derived by using the approximation are 
shown in figure 7, when $\beta$ is changed from 1.0 to 10.0
in two cases of $a_*=0$ and  0.2 with $M=1.4 M_\odot$.
This figure should be compared with figure 5 obtained without the approximation.
The comparison shows that in the approximate case the correlation curve shifts in the upper-right direction
almost along the curve, compared with the case without the approximation. 
In other words, if an observed frequency correlation is described by the results obtained by using
the approximation, we have larger values of $\beta$ compared with those derived without the approximation.
This can be understood from figure 1, where the frequencies calculated by the approximation are shown to be
systematically higher than those calculated without using the approximation.
Comparison of figure 7 with figure 5 also shows that  
the approximation gives systematically a larger $a_*$, if other parameter values are fixed to be the same in
both cases.

\begin{figure}
\begin{center}
    \FigureFile(80mm,80mm){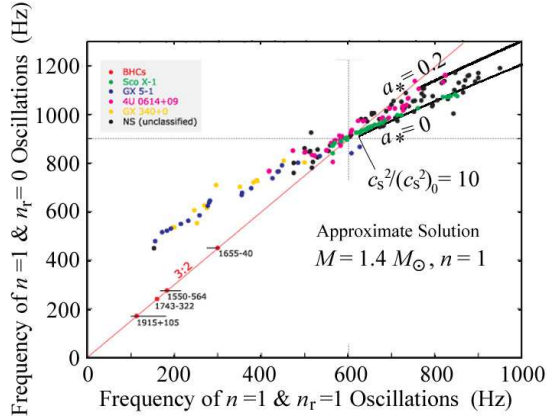}
\end{center}
\caption{The same as figure 5 except that the frequencies are calculated by using the approximation 
that the horizontal motions associated with the oscillations are small perturbations over
the vertical motions.  (Color Online)}
\end{figure}

\subsection{Frequency Correlation between (1,0) and (2,0) Oscillations}

The frequency correlation between two oscillations of $(n, n_{\rm r})=(1,0)$ and $(2,0)$ is
examined in order to to see whether the calculated correlation curve can describe the observational one
between the upper kHz QPO and HBO found by Psaltis et al. (1999).
Two cases of $\eta_{\rm s}=1.3$ and $\eta_{\rm s}=1.1$ are shown in figure 8 by changing 
$\beta$ from 1.0 to 20.0.
The mass of the central source is $M=1.4M_\odot$ and the spin parameter $a_*=0$.
The abscissa is the frequency of the (1,0) oscillation, and the ordinate is that of the (2,0)
oscillation.
The curves are superposed on the frequency-frequency diagram plotting observational data points
(figure 2.9 in a review paper by van der Klis 2004).
The blue curve is for $\eta_{\rm s}=1.3$ and the red one is for $\eta_{\rm s}=1.1$
The left-end of these curves is for $\beta=20.0$ and the right-end is for $\beta=1.0$.
For comparison, the correlation curve between the (1,0) and (2,1) oscillations is also shown
by green curve in the case of $\eta_{\rm s}=1.1$
(in this case the ordinate is the frequency of (2,1) oscillation.)

\begin{figure}
\begin{center}
    \FigureFile(100mm,100mm){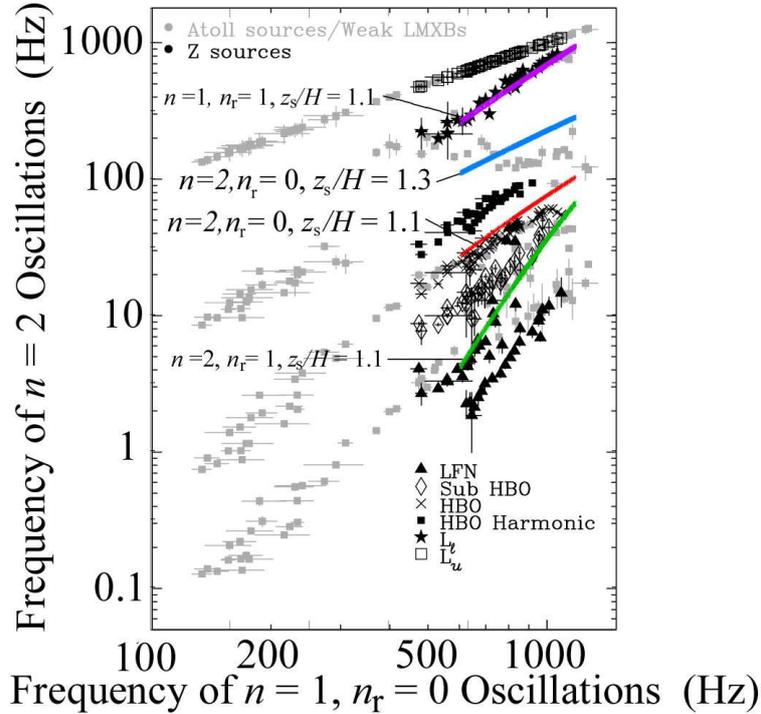}
\end{center}
\caption{The frequency correlation between $(n,\ n_{\rm r})=$ (1,0) and (2,0) oscillations
for two cases of $\eta_{\rm s}=$1.3 (second curve) and 1.1 (third curve).
The value of $\beta$ is changed from 1.0 to 30.0.
The left-end points of the curves are for $\beta=30.0$, and the right-end points are for
$\beta=1.0$.
The curves are overlapped on the plots of observational data which are a part of figure 2.9 by 
van der Klis (2004).
The value of $M$ and $a_*$ adopted are $M=1.4M_\odot$ and $a_*=0$. 
For comparison, the correlation curve between $(n, n_{\rm r})=(1,0)$ and (2,1) is shown (lowermost curve).
The uppermost curve shows, for comparison, the correlation between ($n$, $n_{\rm r})$ $=(1,0)$ and (1,1)
oscillations (i.e., twin kHz QPOs) in the case of $\eta_{\rm s}=1.1$.
(Color Online)}
\end{figure}

We are not interested here in the HBO harmonics and the sub HBO, since their amplitude are small
and will be subsidiary phenomena.
In the case of $\eta_{\rm s}=1.3$ the correlation curve calculated is roughly parallel to the
observed kHz QPO - HBO correlation curve, but runs much above the observed one on the diagram.
As $\eta_{\rm s}$ decreases the curve moves downward on the diagram.
The figure shows that in the case of $\eta_{\rm s}=1.1$ the correlation curve is close to the 
observed sequence of the correlation.
We should notice, however, that in the present paper we have calculated the frequency correlation
of (1,0) and (2,0) oscillations only in a particular case of no magnetic fields.
If disks have strong magnetic fields, 
$\eta_{\rm s}$ required to describe observations will not be so small as $\eta_{\rm s}=1.1$
and also the maximum value of $\beta$ required will become smaller than that shown in figure 8 
(see next section).

\section{Discussion}

Trapping of two-armed ($m=2$) vertical p-mode oscillations in vertically isothermal
disks are examined.
The disks are assumed to be terminated at a certain height by hot, low-density corona.
The height of the termination is taken to be a parameter.

The above issue has been already examined in the previous papers (Kato 2011a,b, 2012a,b) in the 
case where the disks have toroidal magnetic fields, introducing the approximation that
the oscillations are nearly vertical in the sense that the horizontal motions associated with 
the oscillations are small perturbations over the vertical ones.
In the case where disks have no magnetic fields, however, the trapping  
can be examined without the above approximation.
Hence, returning to this simpler case of no magnetic fields, we re-examine in this paper the trapping of 
the vertical p-mode oscillations of $m=2$.
The purposes of this paper are thus i) to examine the case of no magnetic fields more
accuratly without the approximation, and ii) to show that even in the case of no magnetic fields 
the calculated correlation curve can describe the observed one, if disk
parameters such as $\beta\equiv c_{\rm s}^2/(c_{\rm s}^2)_0$ and $\eta_{\rm s}\equiv z_{\rm s}/H$ 
are taken moderately. 

Figure 1 shows that in the case of oscillations of $n=1$, the frequencies calculated without the above-mentioned
approximation are systematically lower compared with those calculated with 
the approximation.
Due to this systematical difference, in the case where the approximation is not used, the correlation curve 
between the (1,0) and (1,1) oscillations 
shifts on the frequency -- frequency diagram in the left-downward direction almost 
along the curve derived with the approximation (compare figures 5 and 7).
We suppose that the same situation occurs even in magnetized disks.
That is, if we could calculate the correlation curves without the approximation in the case of magnetized disks, 
they will shift in the left-downward direction along the curve derived in paper I.
This means that the same correlation range on the diagram can be derived by smaller values of 
$\beta$ and $c_{\rm A}^2/c_{\rm s}$,  
if frequencies coud be calculated without the approximation.
In other words, such strong magnetic fields required in paper I may be unnecessary, 
if we could derive and solve wave equations in magnetized disks without the approximation.
This is in favor of our model, since the magnetic fields required in paper I seem to be somewhat stronger 
than those usually supposed in accretion disks.

Next, we examine whether the (2,0) oscillation can be regarded as the horizontal 
branch oscillation (HBO).
Figure 8 shows that in non-magnetized disks the frequency correlation between the (1,0) and (2,0)
oscillations is close to the observed correlation between the upper kHz QPO and HBO, if $\eta_{\rm s}$
is taken to be as small as $\eta_{\rm s}=1.1$.
This suggests that the disks of horizontal branch phase of Z-sources 
are strongly diminished in their vertical thickness. 

Concerning the disks of the horizontal branch, we considered in paper I two possibilities:
The sequence of horizontal branch is i) a sequence of change of toroidal magnetic
fields (figure 9 of paper I) or ii) a sequence of change of disk temperature with moderate
or strong toroidal magnetic fields (figure 10 of paper I).
The first possibility seems to be less promising, since the correlation curve calculated in paper I has
a sharper gradient on the frequency-frequency diagram, compared with the curve of observations
(see figure 9 of paper I).
This sharp gradient comes from the fact that as the toroidal magnetic fields become strong, the frequency 
of the (2,0) oscillation decreases strongly when $n=2$.\footnote{
This comes from the following fact that the frequency of the fast mode (the fast mode of the three MHD
waves) oscillating in the purely vertical direction is given by
$(\omega-m\Omega)^2=[K_{n,{\rm s}}(c_{\rm s}^2+c_{\rm A}^2)/(c_{\rm s}^2+c_{\rm A}^2/2)+1]
\Omega_\bot^2$ [see equation (47) of Kato (2012a) or equation (27) of Kato (2011a)].
This means that the frequency $\omega$ given by 
$\omega=2\Omega-[K_{n,{\rm s}}(c_{\rm s}^2+c_{\rm A}^2)/(c_{\rm s}^2+c_{\rm A}^2/2)+1]^{1/2}
\Omega_\bot$ decreases, as $(c_{\rm s}^2+c_{\rm A}^2)/(c_{\rm s}^2+c_{\rm A}^2/2)$
increases from 1 to 2 by increase of $c_{\rm A}^2/c_{\rm s}^2$ when $n=2$ ($K_{n,{\rm s}}\geq 1$ 
when $n=2$).
}
Hence, the correlation curve calculated by changing the strength of toroidal magnetic fields
has a sharp gradient on the frequency-frequency diagram.
This situation will not be much changed, even if we could calculate the frequency of trapped 
(2,0) oscillations in magnetized disks without using the approximation of nearly vertical oscillation.

Compared with this, the correlation curve obtained by changing the disk temperature can better describe the
observational correlation curve.
In the limit of no magnetic fields the calculated curve is roughly parallel to the observed correlation 
curve on the frequency -- frequency diagram and moves downward on the diagram 
as $\eta_{\rm s}$ decreases (see figure 8).
If toroidal magnetic fields are present ($c_{\rm A}^2/c_{\rm s}^2$ being taken to be constant during 
temperature change), the correlation curve calculated will move downward furthermore
compared with the case of no magnetic field, since an increase of 
$c_{\rm A}^2/c_{\rm s}^2$ will have a similar effect to a decrease of $\eta_{\rm s}$
[see equation 
$\omega=2\Omega-[K_{n,{\rm s}}(c_{\rm s}^2+c_{\rm A}^2)/(c_{\rm s}^2+c_{\rm A}^2/2)+1]^{1/2}
\Omega_\bot$].\footnote{
A decrease of $\eta_{\rm s}$ increases $K_{n,{\rm s}}$, while an increase of $c_{\rm A}^2/c_{\rm s}^2$
increases $(c_{\rm s}^2+c_{\rm A}^2)/(c_{\rm s}^2+c_{\rm A}^2/2)$.
Hence, both have similar effects on frequency $\omega$.
}
This suggests that the observed correlation curve will be described by $\eta_{\rm s}$ larger than $\eta_{\rm s}=1.1$,
when there are toroidal magnetic fields of moderate strength with $c_{\rm A}^2/c_{\rm s}^2=$ const.
The correlation curve shown in figure 10 of paper I will be also consistent with this argument, since
the correlation curve in the figure will shift in the right-downward direction if
we could calculate the correlation curve without using the approximation of nearly vertical oscillations.

The above argument suggests that the horizontal branch of Z-sources is a sequence of temperature change
in disks whose vertical thickness is strongly diminished.
The direction of temperature increase is from right to left on the horizontal branch, since the direction of 
decrease of observed QPO frequencies is from right to left on the branch.
The temperature change will be a result of a change of mass accretion rate.
Thus, this disk model of the Z-branch is similar to and support the disk model proposed by Church et al. (2006)
and Bolucinska-Church et al. (2010) from spectral analyses, although the origins of QPOs are different.

One of possible causes why the disks of horizontal branch are vertically thin will be strong radiation 
from the central source resulting from high mass accretion and disk evaporation toward the corona,
as Church et al. (2006) and Bolucinska-Church et al. (2010) argue.
If moderately strong toroidal magnetic fields are present in disks, it will be in favor of our model of QPOs, 
although only the limiting case of no magnetic fields is examined in this paper.
It is noted here that disks with strong magnetic fields (the plasma $\beta$ being less than unity
and as small as 0.1)
are expected at a stage of transition from optically thin flows (ADAFs) to optically thick disks by
increase of mass accretion rate (Machida et al. 2006, Oda et al. 2007, 2009, 2010),
although the horizontal branch of Z-sources is different from such stage.

It should be emphasized that in our model of kHz QPOs the correlation curve between the (1,0) and (1,1)
oscillations moves almost along a common curve on the frequency - frequency diagram for changes of
disk parameters, as far as the mass and spin of the central source are fixed.
On the other hand, the correlation curve between the (1,0) and (2,0) oscillations are rather sensitive
to the disk parameters such as $\eta_{\rm s}$.
The former is in favor of our model, since the common curve mentioned above can well describe the
observed correlation curve for reasonable mass and spin parameter of the central source.
However, the latter has two sides for and against our model.
If our model is correct, it is a good tool to determine the disk structure of horizontal branch
of Z-sources.
It is, however, unclear why disk thickness (and magnetic fields) in real disks is really adjusted  to
values relevant to describe observations.

In our model the radius of the inner edge of disks is assumed to be unchanged during
the disk evolution along the sequence.
It is unclear at the present stage of our model whether time change of the inner edge of disks
is required  to describe the observed correlation curve by our model.

The observed amplitude of HBOs is much larger than that of kHz QPOs.
Whether this can be described by our model is a subject to be discussed further.
One of possibilities will be the difference of width of the trapped regions.
In our model, the trapped region of the (1,0) oscillation is only in the innermost region, 
but that of the (2,0) oscillation is wide when $\eta_{\rm s}$ is small (see figure 3).
This difference will be one of causes of amplitude difference between kHz QPOs and HBOs.
It is noted that the trapped region of the (1,1) oscillation is also wide (see figure 2), but in this case
the oscillation has one wavelength in the trapped region.
Hence, by a cancellation by phase mixing the observed amplitude will not become large.
It is also noted that a wide trapped region implies that our basic approximation that $H$ and $\beta$ are
constant in the trapped region will not always be relevant.
Radial variations of $H$ and $\beta$ should be taken into account to do more detailed comparison 
with observations.

\bigskip\noindent
{\bf Appendix A Relation between Trapping Conditions with and without Approximation of 
Nearly Vertical Oscillations}

In a previous paper (Kato 2012a), we have examined the effects of toroidal magnetic fields
on frequencies of trapped oscillations in truncated disks.
In the study we have introduced an approximation that the vertical p-mode oscillations consist of 
nearly vertical motions and thus 
the horizontal motions associated with them can be taken to be small perturbations over the vertical ones.
Under this approximation, in addition to the WKB approximation, we obtained an equation describing
the frequencies of trapped oscillations [see equations (39), (41), and (44) in Kato (2012a)]
\begin{equation}
        \int_{r_i}^{r_{\rm c}}\frac{[(\omega-m\Omega)^2-\kappa^2]^{1/2}}{\Omega_\bot H}
           \frac{\epsilon^{1/2}}{A_{n,{\rm s}}^{1/2}}dr
            =\pi \biggr(n_{\rm r}+\frac{3}{4}\biggr),
\label{A1}
\end{equation}
where $A_{n,{\rm s}}$ is a dimensionless quantity given by equation (37) in Kato (2012a).
Furthermore, $\epsilon$ is a dimensionless quantity defined by [see equations (12) and (17) in Kato (2012a)]
\begin{equation}
    \epsilon(r)=\frac{(\omega-m\Omega)^2-\Omega_\bot^2}{c_{\rm s}^2+c_{\rm A}^2}H^2-K_{n,{\rm s}},
\label{A2}
\end{equation}
and is assumed to be $\epsilon < 1$ (Kato 2012a).

Since in the case of no magnetic fields ($c_{\rm A}^2=0$), $A_{n,{\rm s}}$ is reduced to
\begin{equation}
     A_{n,{\rm s}}=K_{n,{\rm s}}+1={\tilde K}_{n,{\rm s}},
\label{A3}
\end{equation}
equation (\ref{A1}) can be written as
\begin{equation}
    \int_{r_{\rm i}}^{r_{\rm c}}
          \frac{[(\omega-m\Omega)^2-\kappa^2]^{1/2}}{c_{\rm s}}
          \biggr[\frac{(\omega-m\Omega)^2-{\tilde K}_{n,{\rm s}}\Omega_\bot^2}
          {{\tilde K}_{n,{\rm s}}\Omega_\bot^2}\biggr]^{1/2}dr
          =\pi\biggr(n_{\rm r}+\frac{3}{4}\biggr).
\label{A4}
\end{equation}
This is the equation corresponding to equation (\ref{3.8}) in this paper, which is 
\begin{equation}
    \int_{r_{\rm i}}^{r_{\rm c}}
          \frac{[(\omega-m\Omega)^2-\kappa^2]^{1/2}}{c_{\rm s}}
    \biggr[\frac{(\omega-m\Omega)^2-{\tilde K}_{n,{\rm s}}\Omega_\bot^2}
          {(\omega-m\Omega)^2}\biggr]^{1/2}dr
          =\pi\biggr(n_{\rm r}+\frac{3}{4}\biggr).
\label{A5}
\end{equation}
This means that the perturbation method adopted by Kato (2012a) is valid only when
in the propagation region of oscillations we have
\begin{equation}
    (\omega-m\Omega)^2\sim {\tilde K}_{n,{\rm s}}\Omega_\bot^2
\label{A6}
\end{equation}
or 
\begin{equation}
     \omega\sim m\Omega-{\tilde K}_{n,{\rm s}}^{1/2}\Omega_\bot.
\label{A7}
\end{equation}
This is nothing but the condition of $\epsilon < 1$, as is expected.

In the case of two armed oscillations, $m=2$, with $n=1$, we have ${\tilde K}_{n,{\rm s}}=1$.
Hence, this condition is realized  when $\omega$ is comparable with $\Omega$.
In oscillations with $n=2$, ${\tilde K}_{n,{\rm s}}=2$ in  disks with $\eta_{\rm s}=\infty$
and ${\tilde K}_{n,{\rm s}}>2$ in the disks terminated in the vertical thickness (see table 1).
Hence, this approximation is valid when $\omega$ is moderately smaller than $\Omega$ as
is shown in figure 1.

\bigskip
\leftskip=20pt
\parindent=-20pt
\par
{\bf References}
\par
Abramowicz, M. A. 2005, Astron. Nachr. 326, 782   \par
Ba\l ci\'{n}ska-Church, M., Gibiec, A., Jacson, N.K., \& Church, M., 2010, A\& A, 512, A9\par
Church, M.J., Halai, G.S., \& Ba\l ci\'{n}ska-Church, M. 2006, A\&A, 460, 233 \par
Fu, W., \& Lai, D. 2009, ApJ, 690,1386 \par
Kato, S. 2001, PASJ, 53, 1\par 
Kato, S. 2010, PASJ, 62, 635 \par
Kato, S. 2011a, PASJ, 63, 125 \par
Kato, S. 2011b, PASJ, 63, 861 \par
Kato, S. 2012a, PASJ, 64, in press \par
Kato, S. 2012b, PASJ, 64, in press (paper I) \par
Kato, S., Fukue, J., \& Mineshige, S. 2008, Black-Hole Accretion Disks --- Towards a New paradigm --- 
  (Kyoto: Kyoto University Press), chaps. 3, 11 \par
Machida, M., Nakamura, K.E., \& Matsumoto, R., 2006, PASJ, 58, 193 \par
Oda, H., Machida, M., Nakamura, K.E., Matsumoto, R. 2007, PASJ 59, 457 \par 
Oda, H., Machida, M., Nakamura, K.E., Matsumoto, R. 2009, ApJ, 697, 16 \par 
Oda, H., Machida, M., Nakamura, K.E., Matsumoto, R. 2010, ApJ, 712, 639 \par 
Okazaki, A.T., Kato, S., \& Fukue, J. 1987, PASJ, 39, 457 \par
Ortega-Rodrigues, M., Silbergleit, A.S., Wagoner, R. 2008, Geophys. \& Astrophys. Fluid Dynamics,
     102, 75 \par
Lai, D., \& Tsang, D. 2009, MNRAS, 393, 979\par
Psaltis, D., Belloni, T., \& van der Klis, M. 1999, ApJ, 520, 262 \par
Silbergleit, A.S., Wagoner, R., \& Ortega-Rodriguez, M. 2001, ApJ, 548, 335\par
Wagoner, R.V. 1999, Physics Reports, 311, 259 \par
van der Klis, M. 2004, in Compact stellar X-ray sources (Cambridge University Press), eds.
   W.H.G. Lewin, \& M. van der Klis (astro-ph/0410551)
\bigskip\bigskip

\end{document}